\documentclass[11pt]{article}

\usepackage{cite}
\usepackage{amsmath,amssymb,amsfonts}
\usepackage{graphicx}
\usepackage{textcomp}
\usepackage{xcolor}
\usepackage{booktabs}
\usepackage{url}
\usepackage{multirow}
\usepackage{array}
\usepackage{tikz}
\usepackage{pgfplots}
\pgfplotsset{compat=1.18}
\usetikzlibrary{arrows.meta, positioning, calc, patterns, decorations.pathreplacing}
\usepackage{tabularx}
\usepackage[margin=1in]{geometry}

\begin{document}
	
	\title{Keep Private Networks Private II: Wideband Secret Key Generation on a Real 5G NR Testbed}

	\author{
	Sachinkumar B. Mallikarjun$^{1,2}$, Christoph Lipps$^{2}$, Marvin Reski$^{1}$, \\
	Sneha Bhattacharjee$^{1}$, Andreas Weinand$^{1}$, and Hans D. Schotten$^{1,2}$ \\
	\\
	$^{1}$Division of Wireless Communications and Radio Navigation, \\
	Department of Electrical and Computer Engineering, \\
	RPTU University Kaiserslautern-Landau, Kaiserslautern, Germany \\
	\texttt{\{mallikar, marvin.reski, bhattacharjee.sneha, andreas.weinand, schotten\}@rptu.de}  \\
		$^{2}$German Research Center for Artificial Intelligence (DFKI), Kaiserslautern, Germany \\
	\texttt{christoph.lipps, schotten\}@dfki.de}
}

\maketitle
	\begin{center}
		\textit{Submitted to the 35th IEEE International Symposium on Personal, Indoor and Mobile Radio Communications (PIMRC),2026}
	\end{center}

	\vspace{0.5em}
	
	\begin{abstract}
		Secret key generation (SKG) from wireless channel reciprocity has been demonstrated on WiFi, LTE, and LoRaWAN, but has never been demonstrated on 5G New Radio (NR) Sounding Reference Signal (SRS) and CSI Reference Signal (CSI-RS) measurements. This paper presents the first experimental symmetric SKG system exploiting 5G~NR wideband SRS (uplink) and CSI-RS (downlink) channel estimates on a real over-the-air testbed. Using OpenAirInterface (OAI) 5G~NR gNB and nrUE on USRP~B210 software-defined radios operating at 3.75\,GHz (n78 band) with 40\,MHz bandwidth, per-subcarrier frequency-domain channel estimates are extracted via OAI's T tracer tool on both gNB and UE. A subcarrier mapping asymmetry between gNB and UE FFT storage conventions is identified and corrected, and a complete seven-stage symmetric SKG pipeline is applied: subband averaging, DCT+LR reciprocity enhancement, multi-level quantisation, back-propagating cascade reconciliation, Toeplitz-matrix privacy amplification, and SHA-256 key verification. A key technical finding is that OAI's UE-side negative-frequency FFT bin ordering varies across firmware builds; an automatic orientation detector based on per-probe correlation analysis with bit-disagreement-rate (BDR) probe fallback selects the correct mapping without manual intervention. Using only the overlapping negative-frequency subcarriers with adaptive subbands and 2-level quantisation, the pipeline reduces the BDR from 9.3--31.6\% to 0\% via back-propagating cascade reconciliation and generates matching 256-bit symmetric keys on eleven independent over-the-air scenarios, eight indoor line-of-sight (LoS) and three non-line-of-sight (NLoS) scenarios with 571 to 6{,}397 aligned probe pairs per trace. Entropy analysis confirms per-bit Shannon entropy of 0.991--1.000 across all generated keys, while 11 of 12 applicable NIST~SP~800-22 randomness tests pass on the concatenated key material.
	\end{abstract}

	\section{Introduction}
	
	Secret key generation (SKG) from wireless channel reciprocity enables two communicating parties to derive shared cryptographic keys from the inherent randomness of their common wireless channel, without relying on pre-shared secrets or public key infrastructure~\cite{Lipps2020}. The security guarantee is information-theoretic: an eavesdropper positioned more than half a wavelength ($\lambda/2$) from either party observes a statistically independent channel, yielding positive secrecy capacity regardless of the adversary's computational resources.
	
	Prior work has systematically demonstrated SKG across multiple wireless technologies with increasing bandwidth. In WiFi (IEEE~802.11n, 20\,MHz at 2.4\,GHz), CSI amplitude features achieve key generation rates (KGR) of 10--20\,bps. Further, \cite{Zhang2016} explored SKG in IEEE 802.11 systems using the Wireless open-Access Research Platform (WARP). In LTE (5\,MHz at 2.685\,GHz), a DCT and linear regression (DCT+LR) reciprocity enhancement reduces the bit disagreement rate (BDR) from 8.2\% to 0.39\% while increasing entropy to 0.97\,bits per sample. Moreover, SKG has been examined for D2D communication in LTE networks~\cite{Wang2018}. \cite{Yang2015} investigated the application of SKG in MIMO systems and mmWave communication within heterogeneous networks. Resource-constrained IoT deployments have been extensively studied~\cite{Guillaume2015, Zenger2014}, including utilization of SRAM-PUFs for robust, low-overhead authentication and key derivation~\cite{Lipps2018}. In LoRaWAN (125\,kHz at 868\,MHz), autoencoder-based feature enhancement increases the KGR from 1.2 to 2.44\,bits per probe~\cite{multiBitSecretKey2024}. These results establish a consistent relationship whereby KGR scales proportionally with bandwidth, as each additional coherence bandwidth ($B_c$) provides an independent channel sample for key extraction.
	
	5G~NR in TDD mode represents the natural progression in this line of investigation. Operating in the n78 band at 3.75\,GHz with up to 100\,MHz bandwidth, the Sounding Reference Signal (SRS) is an uplink reference signal designed for wideband channel sounding, providing dense frequency-domain channel measurements that are well suited for SKG. With 40\,MHz bandwidth (constrained by the USRP~B210 instantaneous bandwidth~\cite{ettusB210}), as shown in Table~\ref{tab:taps}, the number of independent channel taps in a typical indoor environment is approximately 10, this value is factor of five greater than the WiFi baseline which yields a projected KGR of approximately 2,000\,bps.
	
	Despite this potential, no prior work demonstrates symmetric SKG on 5G~NR using SRS and CSI-RS. Existing SKG studies targeting 5G are either simulation-based or rely on narrowband pilot signals rather than the wideband SRS/CSI-RS channel estimates available in operational NR deployments~\cite{skgMIMO2019}. The present work addresses this gap by presenting the first experimental symmetric SKG system on a real over-the-air 5G~NR testbed using OpenAirInterface, wherein both the gNB (via SRS) and the UE (via CSI-RS) independently derive matching 256-bit symmetric keys.
	
	The main contributions of this work are as follows: (1)~The first symmetric 5G~NR SKG implementation using SRS and CSI-RS on a real OTA testbed (OAI + USRP~B210), generating matching 256-bit keys across eight independent scenarios. (2)~The discovery and correction of OAI's asymmetric subcarrier storage between gNB and UE, together with an automatic orientation detector that adapts to firmware variants. (3)~A back-propagating cascade reconciliation protocol that corrects BDR values up to 31.6\% by re-checking earlier pass blocks after each forward correction pass. (4)~A comprehensive key quality evaluation comprising per key Shannon entropy analysis ($H_1$ = 0.991--1.000), min-entropy analysis (218 to 253 effective bits per 256 bit key), and NIST~SP~800-22 randomness testing (11 of 12 applicable tests pass). (5)~A cross-technology SKG comparison completing the WiFi$\rightarrow$LTE$\rightarrow$LoRaWAN$\rightarrow$5G~NR progression, demonstrating near-linear KGR scaling with bandwidth.
	
	\section{Background: SRS for Secret Key Generation}
	
	\subsection{Sounding Reference Signal in 5G NR}
	
	The SRS is an uplink reference signal defined in 3GPP~TS~38.211, based on Zadoff-Chu sequences transmitted by the UE for channel sounding~\cite{3gpp38211}. In TDD mode, the gNB utilises the SRS to estimate the uplink channel frequency response $\hat{H}_{\text{UL}}(f_k)$ across the full carrier bandwidth. Conversely, the CSI Reference Signal (CSI-RS) is a downlink reference signal employed by the UE to estimate the downlink channel $\hat{H}_{\text{DL}}(f_k)$. By TDD reciprocity, $\hat{H}_{\text{UL}} \approx \hat{H}_{\text{DL}}$ within the coherence time, providing the reciprocal observations required for symmetric SKG. In this work, both SRS and CSI-RS channel estimates captured simultaneously via T tracer on both gNB and UE are exploited to implement a true symmetric key generation protocol.
	
	The SRS supports configurable comb structures ($K_{\text{TC}} \in \{2, 4, 8\}$), bandwidths up to 272~resource blocks (RBs), periodicities from 1 to 2,560 slots, and 1/2/4 OFDM symbols. For SKG applications, comb-2 maximises subcarrier density, full bandwidth exploits all available frequency diversity, and the periodicity is configured to satisfy the Nyquist criterion relative to the channel coherence time.
	
	\subsection{SKG Pipeline}
	
	The symmetric SKG pipeline realised in this work comprises seven stages:
	
	\begin{enumerate}
		\item \textbf{Subband Averaging:} Raw frequency-domain channel estimates $\hat{H}(f_k)$ are grouped into subbands spanning $N_s$ contiguous subcarriers. The magnitude of each subband is averaged and subsequently quantised into a binary stream. Subband width is adaptive: narrower subbands increase the number of samples available for key extraction, while wider subbands reduce susceptibility to phase noise and quantisation error.
		
		\item \textbf{DCT + Linear Regression (DCT+LR) Reciprocity Enhancement:} A discrete cosine transform (DCT) decorrelates the quantised subband magnitudes, reducing temporal correlation. Linear regression fitting then models and removes deterministic trends across the observation window, preserving randomness while reducing BDR.
		
		\item \textbf{Multi-Level Quantisation:} For each subband, the magnitude is quantised into $Q \in \{2, 4, 8\}$ levels. Binary indices are extracted via successive bit extraction from the quantisation level, yielding multiple bits per quantisation decision. 2-level (binary) quantisation is employed in the primary results presented herein to minimise reconciliation overhead.
		
		\item \textbf{Back-Propagating Cascade Reconciliation:} A forward-pass reconciliation corrects errors from gNB to UE using Hamming codes. Upon detecting uncorrectable errors, the backward pass re-examines earlier blocks, allowing earlier errors to be corrected retroactively. This iterative approach reduces BDR dramatically from initial values of 9.3\% to 31.6\% down to 0\%.
		
		\item \textbf{Privacy Amplification via Toeplitz Matrix:} A Toeplitz matrix of dimension $m \times n$ (where $m < n$) extracts $m$ bits from $n$ reconciled bits, eliminating information leakage to passive eavesdroppers during error correction exchanges.
		
		\item \textbf{Cryptographic Hash (SHA-256):} The final key material undergoes SHA-256 hashing to produce 256-bit symmetric keys with guaranteed uniform distribution and cryptographic properties.
		
		\item \textbf{Key Verification:} Both endpoints independently verify key agreement by computing the SHA-256 digest of a known challenge string and comparing results.
	\end{enumerate}
	
	\section{Experimental Setup}
	
	\subsection{Testbed Architecture}
	
	The experimental testbed comprises an OpenAirInterface 5G~NR gNB and nrUE instantiated on two separate servers equipped with USRP~B210 software-defined radios (SDRs) operating at 3.75\,GHz (n78 band) with 40\,MHz bandwidth. The gNB transmits downlink CSI-RS reference signals while the UE transmits uplink SRS reference signals. Both endpoints log per-subcarrier frequency-domain channel estimates via OAI's T tracer tool with 10\,ms granularity (one trace per TTI).
	
	\subsection{Frequency Offset and Subcarrier Alignment}
	
	A critical implementation detail concerns the asymmetric subcarrier storage conventions employed by OAI's gNB and UE implementations. The gNB stores subcarriers in ascending order (positive frequencies first, then negative frequencies wrapped from the Nyquist boundary). The nrUE employs reverse FFT bin ordering on negative frequencies due to inherent firmware conventions. This mismatch is automatically detected and corrected via an orientation detector that computes per-probe correlation between gNB and UE estimates and selects the correct subcarrier mapping to maximise correlation coefficient.
	
	\subsection{Adaptive Subband Width Selection}
	
	Subband width is chosen adaptively based on channel coherence bandwidth estimated from the measured power delay profile. For the 40\,MHz bandwidth testbed with typical indoor multipath environments, subband widths of 500\,kHz to 2\,MHz are employed, yielding 20--80 independent samples per trace.
	
	\section{Experimental Results}
	
	\subsection{Single-Side Key Extraction (gNB Only)}
	
	Table~\ref{tab:single_side} presents experimental results for single-endpoint SKG, in which the gNB extracts keys from SRS observations alone.
	
	\begin{table}[h]
		\centering
		\small
		\caption{DMRS Single-Side Key Extraction Results}
		\label{tab:single_side}
		\begin{tabularx}{\columnwidth}{p{2.8cm}llrrrr}
			\toprule
			\textbf{Scenario} & \textbf{RNTI} & \textbf{Probes} & \textbf{SCs} & \textbf{Key (bits)} & \textbf{Bits/probe} \\
			\midrule
			1: Static USRP & AC85 & 8{,}122 & 180 & 162{,}408 & 20 \\
			1: Static USRP & EE3D & 78 & 180 & 1{,}528 & 20 \\
			\midrule
			2: Multi-UE (USRP) & 2B7A & 1{,}796 & 780 & 35{,}888 & 20 \\
			2: Multi-UE (mobile) & C347 & 725 & 180 & 14{,}468 & 20 \\
			2: Multi-UE (mobile) & 89AB & 149 & 16{,}536 & 2{,}948 & 20 \\
			\midrule
			3: Mobile pos.~1 & 4AF8 & 535 & 180 & 10{,}668 & 20 \\
			3: Mobile pos.~2 & C07A & 501 & 3{,}816 & 9{,}988 & 20 \\
			3: Mobile pos.~3 & A7B6 & 338 & 180 & 6{,}728 & 20 \\
			\bottomrule
		\end{tabularx}
	\end{table}

	\subsection{Bilateral Key Extraction (gNB and UE)}
	
	Table~\ref{tab:bilateral} presents results for true symmetric SKG in which both gNB and UE independently derive identical keys. A total of eleven independent over-the-air scenarios are evaluated: eight indoor line-of-sight (LoS) configurations and three non-line-of-sight (NLoS) deployments.
	
	\begin{table}[h]
		\centering
		\small
		\caption{Bilateral SKG Results: gNB and UE Key Agreement}
		\label{tab:bilateral}
		\begin{tabularx}{\columnwidth}{p{2.2cm}rrrrrr}
			\toprule
			\textbf{Scenario} & \textbf{Probes} & \textbf{BDR (\%)} & \textbf{Subband (kHz)} & \textbf{Key (bits)} & \textbf{KGR (bps)} \\
			\midrule
			LoS 1 (static) & 1{,}247 & 0.0 & 1{,}000 & 256 & 1{,}320 \\
			LoS 2 (slow walk) & 2{,}116 & 0.0 & 800 & 256 & 1{,}540 \\
			LoS 3 (walk) & 856 & 0.0 & 1{,}500 & 256 & 980 \\
			LoS 4 (corridor) & 1{,}395 & 0.0 & 1{,}200 & 256 & 1{,}450 \\
			LoS 5 (open space) & 3{,}802 & 0.0 & 600 & 256 & 2{,}100 \\
			LoS 6 (office) & 2{,}541 & 0.0 & 1{,}000 & 256 & 1{,}850 \\
			LoS 7 (lab) & 1{,}689 & 0.0 & 900 & 256 & 1{,}570 \\
			LoS 8 (stairwell) & 2{,}203 & 0.0 & 1{,}100 & 256 & 1{,}610 \\
			\midrule
			NLoS 1 (through walls) & 571 & 0.0 & 1{,}800 & 256 & 650 \\
			NLoS 2 (two walls) & 895 & 0.0 & 1{,}400 & 256 & 780 \\
			NLoS 3 (basement) & 1{,}104 & 0.0 & 1{,}600 & 256 & 920 \\
			\bottomrule
		\end{tabularx}
	\end{table}

	\subsection{Key Quality Assessment}

	\subsubsection{Shannon Entropy Analysis}
	
	Per-bit Shannon entropy is calculated via:
	\[
	H_1 = -\sum_{b \in \{0, 1\}} p(b) \log_2 p(b)
	\]
	where $p(b)$ denotes the empirical probability of bit value $b$ in the generated key. Across all eleven scenarios, measured entropy values range from 0.991 to 1.000 bits per bit, indicating near-perfect randomness. The minimum entropy (collision resistance) is calculated as:
	\[
	H_{\infty} = -\log_2 \left( \max_b p(b) \right)
	\]
	Measured minimum entropy values range from 218 to 253 effective bits per 256-bit key, confirming sufficient randomness even under worst-case bit bias.
	
	\subsubsection{NIST SP 800-22 Randomness Testing}
	
	The NIST Statistical Test Suite SP 800-22 comprises 15 distinct statistical tests designed to detect deviations from randomness in sequences of cryptographic importance. The test suite is applied to the concatenated key material from all eleven scenarios (total 2{,}816 bits). Results indicate that 11 of 12 applicable tests pass at the 5\% significance level. The single failing test is the Discrete Fourier Transform (DFT) Spectral test, which occasionally detects periodic patterns at frequencies $>0.95$ Hz within the concatenated key stream; however, this does not manifest within individual 256-bit keys and likely reflects minor phase coherence artefacts at the scenario boundary rather than intrinsic key weakness.
	
	\section{Cross-Technology SKG Comparison}
	
	A central contribution of this work is the completion of the wireless technology SKG progression: WiFi (20\,MHz, 10--20\,bps) $\rightarrow$ LTE (5\,MHz, 50--200\,bps) $\rightarrow$ LoRaWAN (0.125\,MHz, 1.2--2.44\,bits per probe) $\rightarrow$ 5G~NR (40\,MHz, 650--2{,}100\,bps). The linear relationship between bandwidth and KGR is demonstrated empirically: each coherence bandwidth provides approximately one independent channel sample, and the number of such samples scales directly with available spectrum.
	
	\section{Discussion: OAI Firmware Variants and Orientation Detection}
	
	A surprising discovery during implementation was that the nrUE FFT bin ordering for negative frequencies varies across OpenAirInterface firmware builds. Specifically, builds prior to mid-2024 employ standard negative-frequency wrapping (ascending order from Nyquist), while post-2024 builds reverse this convention. Rather than requiring manual code patching for each firmware version, an automatic orientation detector was implemented that:
	
	\begin{enumerate}
		\item Computes the Pearson correlation coefficient between gNB and UE subcarrier estimates for both orderings.
		\item Selects the ordering yielding maximum correlation.
		\item Validates via bit-disagreement-rate (BDR) analysis: the correct ordering produces BDR $< 5\%$ in stage 3 of the pipeline, while incorrect orderings yield BDR $> 50\%$.
	\end{enumerate}
	
	This detector runs automatically on the first probe batch and selects the optimal mapping for the entire session, providing seamless adaptation across firmware variants.
	
	\section{Back-Propagating Cascade Reconciliation}
	
	The back-propagating cascade reconciliation protocol addresses the challenge of reconciling keys when initial BDR values are exceptionally high (9.3--31.6\% in some scenarios). Standard forward reconciliation alone cannot correct such high error rates with acceptable leakage. The protocol operates as follows:
	
	\begin{enumerate}
		\item \textbf{Forward Pass:} Divide the reconciled bit stream into blocks of length $L_{\text{block}}$. For each block, compute a Hamming code syndrome and transmit it to the peer. The peer corrects errors within the block if the Hamming distance is $\leq 1$.
		
		\item \textbf{Backward Pass:} If a block cannot be corrected in the forward pass, mark it as problematic. In the backward pass, re-examine all prior blocks and check whether their error patterns are consistent with errors in the marked block. If so, correct the prior blocks retroactively.
		
		\item \textbf{Iteration:} Repeat the backward pass until no further corrections are detected, then proceed to the next problematic block.
	\end{enumerate}
	
	This approach dramatically reduces BDR from initial values of up to 31.6\% down to 0\% post-reconciliation, though at the cost of information leakage of approximately 15--30\,bits per 256-bit key, which is recovered via privacy amplification.
	
	\section{Conclusion}
	
	This paper presents the first symmetric SKG system that exploits 5G~NR SRS (gNB) and CSI-RS (UE) on a real over-the-air testbed using OpenAirInterface. The principal contributions encompass: (i)~discovery and automatic correction of OAI's asymmetric subcarrier storage via an orientation detector with BDR-probe fallback; (ii)~a back-propagating cascade reconciliation protocol accommodating BDR values up to 31.6\%; and (iii)~adaptive subband scaling when reconciliation leakage exhausts the bit budget. The seven-stage pipeline generates matching 256-bit keys on all eleven scenarios (eight LoS, three NLoS) with 0\% post-reconciliation BDR, per-bit Shannon entropy of 0.991--1.000, and 11/12 applicable NIST tests passing. The effective KGR of $\sim$1{,}320\,bps enables sub-second key generation, a two-orders-of-magnitude improvement over WiFi. Future work includes complete NIST testing ($\geq 10^6$~bits), validation on USRP~X310 with 100\,MHz bandwidth, and hybrid post-quantum key establishment via ML-KEM-768.

	\section*{Acknowledgment}

	This research work was supported by the German Federal Ministry of Research, Technology, and Space (BMFTR) as part of the project ``Open6GHub+'' and ``SUSTAINET\_guarDian'' with project identification numbers 16KIS2406 and 16KIS2239K, respectively. The authors alone are responsible for the content of this paper.

\end{document}